# Exploring Coverage and Distribution of Identifiers on the Scholarly Web

*Peter Kraker[1], Asura Enkhbayar[1], Elisabeth Lex[2]*

[1]Know-Center
Inffeldgasse 13/VI, 8010 Graz
{pkraker, aenkhbayar}@know-center.at

[2]Graz University of Technology
Inffeldgasse 13/V, 8010 Graz
elisabeth.lex@tugraz.at

**Abstract**

In a scientific publishing environment that is increasingly moving online, identifiers of scholarly work are gaining in importance. In this paper, we analysed identifier distribution and coverage of articles from the discipline of quantitative biology using arXiv, Mendeley and CrossRef as data sources. The results show that when retrieving arXiv articles from Mendeley, we were able to find more papers using the DOI than the arXiv ID. This indicates that DOI may be a better identifier with respect to findability. We also find that coverage of articles on Mendeley decreases in the most recent years, whereas the coverage of DOIs does not decrease in the same order of magnitude. This hints at the fact that there is a certain time lag involved, before articles are covered in crowd-sourced services on the scholarly web.

**Keywords**: scholarly identifiers, pre-prints, arXiv, DOI, readership

# Introduction

In a scientific publishing environment that is increasingly moving online, identifiers of scholarly work are gaining in importance. With the advent of pre-print archives, there is often more than one version of an article available and these versions may be hosted in various places around the web. Scholarly communication is no longer limited to articles alone, but it also takes place in different forms on various social media platforms. Identifiers are therefore crucial for disambiguation and traceability of scholarly articles and their reception.

The need for persistent identifiers is often mentioned in the literature (see e.g. Davidson & Douglas, 1998; Bourne and Fink 2008) and consequently, a variety of identifier systems have been proposed (see e.g. Van De Sompel et al., 2001; Warner 2010). Prominent examples for identifiers on an article level are the Digital Object Identifier or DOI (DOI Foundation, n.d.) and the arXiv ID. Notable identifiers on the author level are author-based identifiers such as ORCID (Haak et al., 2012) and Researcher ID (Thomson-Reuters, n.d.). Some of the most longstanding identifiers predate the digital age, including the International Standard Book Number (ISBN) and the International Standard Serial Number (ISSN).

Despite their importance, little is empirically known about the coverage and distribution of scholarly identifiers, and how they propagate on the scholarly web. In our work, we are addressing this very gap in the scientometric literature. Specifically, our research was guided by the following research questions:

- How are scholarly identifiers distributed in crowd-sourced systems, e.g. pre-print archives and online reference management systems? Which identifier combinations are the most common? Who are the top providers of identifiers?
- Does the provision of different identifiers have an influence on findability of scientific publications in other bibliographic and bibliometric sources?

## Data and Method

In this study, we analysed arXiv papers from the discipline of quantitative biology (arXiv short code: q-bio). We chose this discipline because it represents one of the largest disciplines on Mendeley (Kraker et al., 2012). Three different data sources were used in this study: (i) arXiv, a preprint archive (ii)

CrossRef, a metadata and linking service, and (iii) Mendeley, an online reference management system.

The data collection pipeline is shown in Figure 1. At first, we collected metadata on all publicly available articles for quantitative biology. In all cases, the most recent upload to arXiv was used and all older entries were discarded. This resulted in n=14,195 metadata records. Quantitative biology represents a medium-to-small collection on arXiv. The collected metadata includes: arXiv ID, DOI (optional), title, authors, year, and journal (optional).

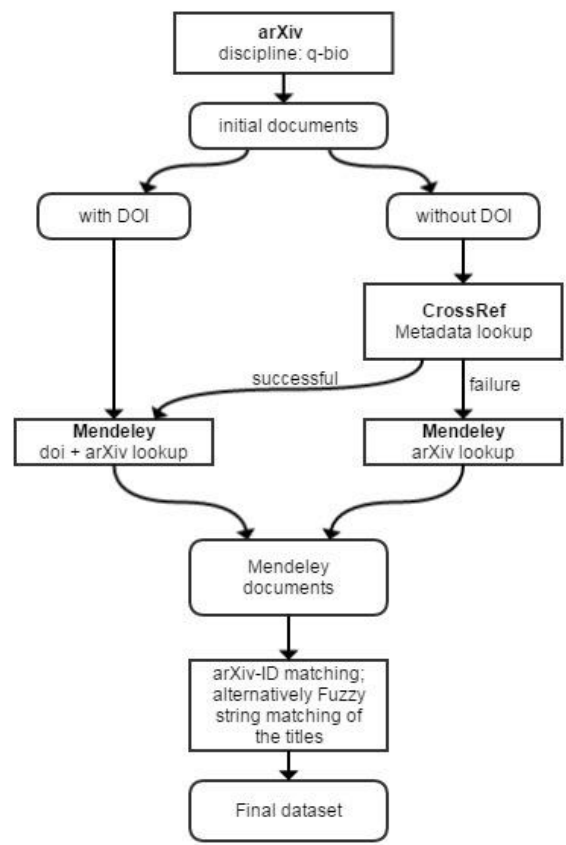

Figure 1: Data collection pipeline

This data was sourced on 17.11.2014 and was used as a basis for all following steps. At first, the initial data set was divided into entries with DOI (n=5,125 entries, 36.7%) and without a DOI (n=8,980 entries, 63.3%). arXiv is primarily used as a way to disseminate pre-prints, and not all authors add a DOI to the arXiv record after an article has been published. Therefore, we performed a CrossRef meta-data lookup in order to acquire additional DOIs. We used the following metadata to search for an entry: title, author, journal, and year.

With this procedure, we found DOIs for an additional 1,885 entries, bringing the number of entries with a DOI up to 7,100 (50.02%). We then attempted to retrieve the corresponding documents for all entries on Mendeley. We used either the arXiv ID or both the DOI and the arXiv ID to locate the document. If both arXiv ID and DOI yielded a result on Mendeley, the Mendeley IDs were compared. If they didn't match, we used the result, which contained additional identifier fields, e.g. a PubMed ID, if available. If both results contained the same amount of articles, we chose the item found with the DOI.

Finally, we compared the arXiv ID of the obtained Mendeley document with the original arXiv entry. If the obtained Mendeley document did not provide one, the two titles were compared using approximate string matching in order to ascertain matching documents.

After this procedure, we arrived a final set of n=11,570 articles that could be found on Mendeley (81.5%). For these articles, we retrieved basic readership data and identifier data. Available identifiers on Mendeley are[1]:

- arxiv: arXiv ID
- doi: Digital Object Identifier (DOI)
- isbn: International Standard Book Number (ISBN)
- issn: International Standard Serial Number (ISSN)
- pmid: PubMed ID (assigned to publications indexed in PubMed)
- scopus: Scopus ID (assigned to publications indexed in Scopus)
- ssrn: Social Science Research Network (SSRN) ID

---

[1] see http://dev.mendeley.com/methods/#catalog-documents

# Results

## Identifier distribution in arXiv and findability on Mendeley

Table 1 sums up the basic results of the crawling process. Of the 14,195 unique articles, 36.7% had a DOI on arXiv. Using CrossRef, an additional 1,885 DOIs could be found, bringing the share of articles with a DOI up to 50.02%. 11,570 articles (81.5%) could finally be found on Mendeley.

There was a difference in findability with respect to whether we used a DOI or the arXiv ID to search for the articles on Mendeley (see also Table 3). Of the 14,195 articles, 72.6% could be retrieved on Mendeley using the arXiv ID. In contrast to that, 91.4% of the 7,100 articles with a Digital Object Identifier (either on arXiv or via metadata lookup on CrossRef) could be found on Mendeley using the DOI.

One of the reasons for that could be that records with a DOI do represent articles that have eventually been published in a journal. In order to test this assumption, we analysed the registrants for all entries with a DOI (7,100 articles). We used a list of DOI registrants by Alf Eaton[2] with manual extensions to identify registrants. The results confirm our assumption (see Table 2). The top registrants are established publishers such as Elsevier and Springer. These publishers usually assign DOIs to articles published in their journals and books, in contrast to archives such as figshare, which assign a DOI to any submitted article regardless of whether it was published in a journal or not.

Table 1: Results of the crawling process; n=14,195 articles

| arXiv: total docs | arXiv: docs with DOI | CrossRef: additional DOIs | Mendeley: found |
|---|---|---|---|
| 14,195 | 5,125 (36.7%) | 1,885 (13.3%) | 11,570 (81.5%) |

---

[2] see https://gist.github.com/hubgit/5974843

Table 2: DOI registrants of articles; n=7.100 articles

| Registrant | # DOIs | Percentage |
|---|---:|---:|
| American Physical Society | 1,507 | 21.2% |
| Elsevier | 1,029 | 14.5% |
| Springer-Verlag | 668 | 9.4% |
| Public Library of Science | 502 | 7.1% |
| IOP Publishing | 439 | 6.2% |
| American Institute of Physics | 335 | 4.7% |
| Proceedings of the National Academy of Sciences | 217 | 3.1% |
| Oxford University Press | 194 | 2.7% |
| Springer (Biomed Central Ltd.) | 180 | 2.5% |
| IOP Publishing - Europhysics Letters | 141 | 2.0% |
| Other | 1,888 | 26.6% |
| **Sum** | **7,100** | **100%** |

To eliminate effects that relate to the nature of the article that has been posted on arXiv (whether it stayed a pre-print or went on to become a journal article), we also compared findability for articles that have both a DOI and an arXiv ID (see Table 3). We also found a difference in these cases: 91.4% of articles with a DOI could be found using the very same identifier, whereas, only 71.4% of articles with a DOI could be found with the arXiv ID. The lowest findability was reported for articles with no DOI: of the 7,095 articles with no DOI, only 69.0% were retrieved using the arXiv ID.

Table 3: Findability of articles on Mendeley, depending on the identifier used; n=14,195 articles

| | n | found on Mendeley using | |
|---|---|---|---|
| | | arXiv ID | DOI |
| **arXiv ID & DOI** | 7,100 (50.02%) | 5,414 (76.25%) | 6,492 (91.44%) |
| **arXiv ID** | 7,095 (49.98%) | 4,896 (69.01%) | - |
| **Sum** | 14,195 (100%) | 10,310 (72.63%) | - |

Another interesting fact found in the top providers is that the American Physical Society, which is, among other things, "working to advance and diffuse the knowledge of physics through its outstanding research journals"[3] is the top registrant for DOIs in quantitative biology. One of the reasons for that could be that arXiv allows authors to assign more than just one category to each article. The analysis of article categories (see Table 4) shows that quantitative biology is the primary discipline for only 61.4% of articles with a DOI (4,358 articles). 30.1% (2,178 articles) are assigned to a primary category that falls into the discipline of physics. This indicates a high number of interdisciplinary articles in the sample.

Figure 2 shows the distribution of articles from 1992 to 2013. There is a strong, at times exponential increase in the number of articles. The coverage on Mendeley, however, has declined for the youngest articles as can be seen in Figure 3. The percentage of articles with a DOI does not decrease in the same order of magnitude.

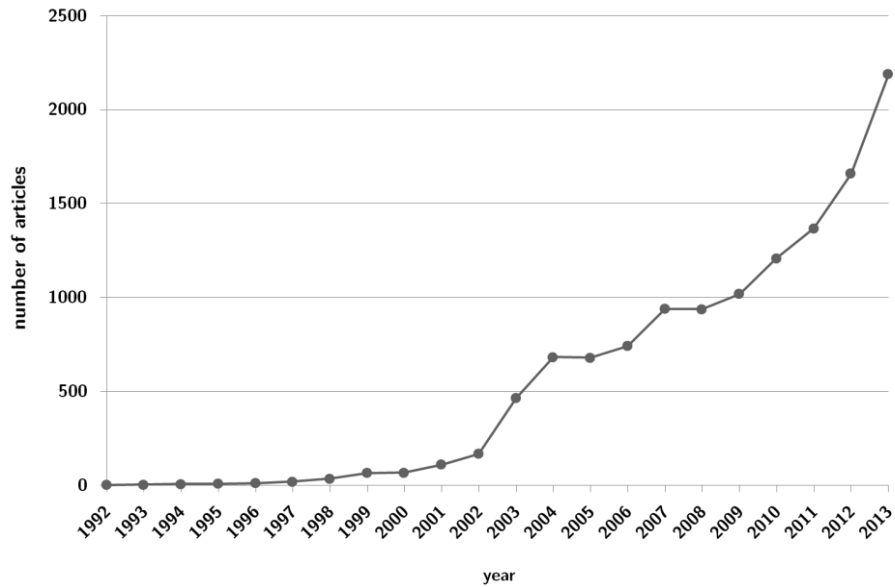

Figure 2: Distribution of articles between 1992 and 2013; n=12,392 articles

---
[3] see http://www.aps.org/about/index.cfm

Table 4: Distribution of disciplines in articles with a DOI (n=7,100 articles)

| Discipline | Number of articles | Percentage |
|---|---|---|
| Quantitative Biology | 4,358 | 61.4% |
| Physics | 2,178 | 30.7% |
| Computer Science | 247 | 3.5% |
| Mathematics | 211 | 3.0% |
| Statistics | 105 | 1.5% |
| Quantitative Finance | 1 | 0.0% |
| **All** | **7,100** | **100.0%** |

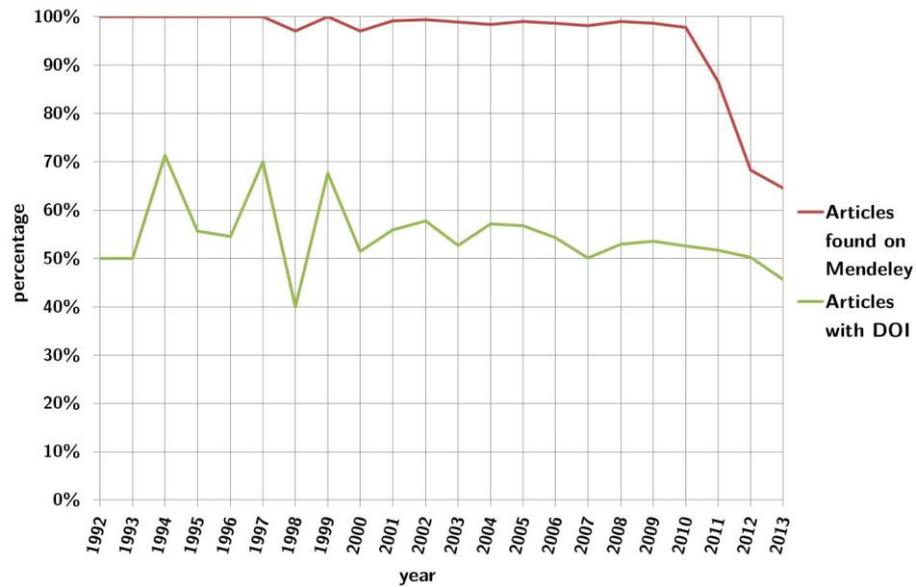

Figure 3: Findability of articles on Mendeley and DOI coverage, 1992-2013; n=12,392 articles

## Distribution of identifiers on Mendeley

We then investigated the distribution of identifiers of all arXiv articles found on Mendeley in detail. Note that we only took metadata from Mendeley into account, which is why the numbers for arXiv ID and DOI differ to the anal-yses before. The distribution of identifiers on Mendeley can be seen in Table 5. The arXiv ID is the most common identifier, followed by the Scopus ID, DOI and ISSN. In terms of readership, articles with a PubMed ID have the highest average readership[4].

Figure 4 shows the most common identifier combinations in the data. Here, a combination of all identifiers on Mendeley included in this analysis (arXiv ID, DOI, ISSN, PubMed ID and Scopus ID) is the most common identifier combination; a single arXiv ID comes second.

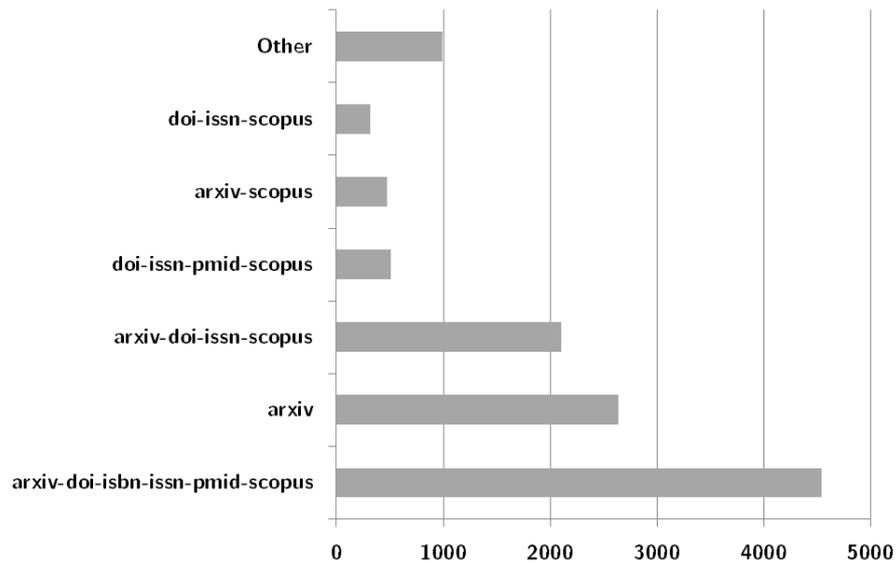

Figure 4: Identifier combination frequency of articles on Mendeley; n=11,570 articles

---

[4] Note that we left ISBN out of this analysis, because the metadata quality was very poor with respect to this field on Mendeley.

Table 5: Identifier frequency and mean readership on Mendeley; n=11,570 articles

|  | arxiv | doi | scopus | pmid | issn |
|---|---|---|---|---|---|
| **frequency** | 10,351 (89.5%) | 8,321 (71.9%) | 8,409 (72.7%) | 5,477 (47.3%) | 8,119 (70.2%) |
| **mean readership** | 20.4 | 25.4 | 25.4 | 32.4 | 25.9 |

.

## Conclusions and Future Work

We found that when retrieving arXiv articles in quantitative biology from Mendeley, we were able to obtain more articles using the DOI than the arXiv ID. Even when we only considered articles that were assigned both identifiers, the effect was sizeable (91.4% vs. 72.6%). This indicates that the DOI may be a better identifier with respect to findability. Nevertheless, a single arXiv ID is the second most popular identifier combination on Mendeley. This suggests that pre-prints are being read – if at a lower level – even when they are not yet published in a journal.

We found that coverage of articles on Mendeley decreases in the most recent years, whereas the availability of DOIs does not decrease in the same order of magnitude. This hints at the fact that there is a certain time lag before articles are covered in crowd-sourced services on the scholarly web.

There are certain limitations to this work. We only looked at a single discipline (quantitative biology) and we only used three data sources in our study (arXiv, CrossRef and Mendeley), which may have had a significant influence on the results. Indeed, in a small-scale study using a random sample of 381 articles from Web of Science, Zahedi et al. (2014) report that they were able to retrieve only 47.7% of articles on Mendeley using the DOI or the title.

In the future, we therefore plan to extend this study to more disciplines and fields in order to substantiate the hypotheses emanating from the results in this study. In order to gain a deeper insight into the distribution and the coverage of identifiers on the scientific web, we are looking to include further data sources such as Web of Science, PubMed Central, Altmetric.com, and figshare.

**Acknowledgments**

The Know-Center is funded within the Austrian COMET program – Competence Centers for Excellent Technologies - under the auspices of the Austrian Federal Ministry of Transport, Innovation and Technology, the Austrian Federal Ministry of Economy, Family and Youth, and the State of Styria. COMET is managed by the Austrian Research Promotion Agency FFG.

**Curriculum Vitae**

Peter Kraker is a postdoctoral researcher in the field of social computing at the Know-Center of Graz University of Technology. He completed his PhD thesis on visualizing research fields based on scholarly communication on the web at University of Graz with honours. His main research interests are Science 2.0, Open Science and Altmetrics. Peter contributed to several EU-funded projects in these areas, including TEAM and STELLAR.

Asura Enkhbayar is junior researcher in the field of social computing at the Know-Center of Graz University of Technology. He received his bachelor's degree in Electronic Engineering from the University of Applied Sciences Technikum Vienna. His main research interests are Scholarly Communication, Data Mining and Analysis, Machine Learning, Medical Image Processing and Data Journalism.

Elisabeth Lex is the head of the Social Computing group at Know-Center and she has an assistant professor position at Graz University of Technology. Her research interests include Social Computing, Science 2.0, Web Science, Social Network Analysis, Machine Learning, Information Retrieval, and Data Mining. Elisabeth is work package leader in the FP7 IP Learning Layers and coordinates the FP7-PEOPLE-2011-IRSES project WIQ-EI.